\def \<{\langle}
\def \>{\rangle}

\documentclass[a4paper,11pt]{article}  %>>> use for US letter paper
\usepackage{jcappub,graphicx,epsfig,natbib,color,times,bm,amsmath,multirow,hyperref}

\usepackage{subfig}
\usepackage{threeparttable}
\usepackage{booktabs}

\newcommand{\tred}{\textcolor{red}}

\title{A Constrained NILC method for CMB B mode observations }

\author[a]{Zirui Zhang}\emailAdd{zhzr@mail.sdu.edu.cn}
\author[b]{Yang Liu\footnote{Corresponding author\label{corresauth}}}\emailAdd{liuy92@ihep.ac.cn}
\author[b]{Si-Yu Li}\emailAdd{lisy@ihep.ac.cn}
\author[a]{Haifeng Li}\emailAdd{lihf@sdu.edu.cn}
\author[b]{Hong Li\textsuperscript{\ref{corresauth}}}\emailAdd{hongli@ihep.ac.cn}
\affiliation[a]{Institute of Frontier and Interdisciplinary Science and Key Laboratory of Particle Physics and Particle Irradiation (MOE), Shandong University, Qingdao 266237, China}
\affiliation[b]{Key Laboratory of Particle Astrophysics, Institute of High Energy Physics, Chinese Academy of Sciences, Beijing 100086, China}

\affiliation[c]{University of Chinese Academy of Sciences, Beijing 100049, China}

\begin{document}

\abstract{
The Internal Linear Combination (ILC) method is commonly employed to extract the cosmic microwave background (CMB) signal from multi-frequency observation maps. 
However, the performance of the ILC method tends to degrade when the signal-to-noise ratio (SNR) is relatively low, particularly when measuring the primordial $B$-modes to detect the primordial gravitational waves.
%\tred{\sout{To address this issue, we propose an enhanced version of needlet ILC (NILC) method on $B$ map called constrained NILC which is more suitable for situations with low SNR by adding additional prior foreground information.} }
To address this issue, an enhanced version of the ILC method, known as constrained ILC, is proposed.
This method is designed to be more suitable for situations with low signal-to-noise ratio (SNR) by incorporating additional prior foreground information. 
In our study, we have modified the constraint Needlet ILC method and successfully improved its performance at low SNR.
We illustrate our methods using mock data generated from the combination of WMAP, Planck and a ground-based experiment in the northern hemisphere, and the chosen noise level for the ground-based experiment are very conservative which can be easily achieved in the very near future. The results show that the level of foreground residual can be well controlled. In comparison to the standard NILC method, which introduces a bias to the tensor-to-scalar ratio ($r$) of approximately $0.05$, the constrained NILC method exhibits a significantly reduced bias of only around $5\times10^{-3}$ towards $r$ which is much smaller than the statistical error.

%\tred{The goal:}To develop an efficient method for CMB component seperation, the so called constrained NILC method in B map. %We consider both ground based CMB observation, a partial sky map case, and a future space mission case, a full sky map case. 

%\tred{The way:}With simulations, we illustrate the detail techniques for the methods, how to handle with the point sources, how to optimise the mask, with the simulations of AliCPT-like experiments. 

%\tred{The conclusion: Compared with the previous study, we generate improvement in the aspects of ....}
}

%\keywords{methods: data analysis – CMB polarization $B$ mode}
\keywords{CMB Polarization, Foreground Removal, ILC, Primordial Gravitational Waves}
\maketitle

\section{Introduction}\label{sec:intro}
The cosmic microwave background (CMB) radiation, originating approximately 380,000 years after the Big Bang, represents the oldest light in the universe and carries a wealth of cosmological information. Over the past decades, significant advancements in CMB observations have made it the most accurate cosmic probe, playing a pivotal role in establishing the standard cosmological model known as $\Lambda$CDM, and ushering in a new era of precise cosmology.
% }

%\textcolor{blue}{
To address certain challenges encountered within the $\Lambda $CDM model, such as the horizon, flatness, and magnetic monopole problems, the inflation theory was proposed\cite{STAROBINSKY1982175, PhysRevD.23.347}. In addition to resolving these issues, inflation theory also predicted the existence of primordial gravitational waves (PGWs). 
The detection of PGWs stands as a crucial frontier in cosmology, with the precise measurement of the CMB's primordial $B$-mode polarization signal serving as the most promising approach. The strength of this signal is directly proportional to the amplitude of PGWs, which can be quantified by the tensor-to-scalar ratio $r$. Various experimental groups, such as BICEP/Keck series \cite{Bicep2014, BK2015, BicepKeck:2021ybl}, Simons Observatory\cite{SimonsObservatory:2018koc}, CMB-S4\cite{S419}, LiteBIRD \cite{2019JLTP..194..443H} etc., are actively involved in the pursuit of detecting and characterizing PGWs.%}

% \textcolor{blue}{
Currently, the strongest constraint on $r$ was given from the joint analysis of BICEP2/\textit{Keck Array}, Planck and WMAP as $r_{0.05} < 0.036$ at $95\%$ confidence level\cite{2021PhRvL.127o1301A}. However, detecting and characterizing the primordial $B$-mode signal is a formidable challenge due to its inherent weakness compared to other sky signals. One of the foremost obstacles in primordial $B$-mode measurements is the presence of foreground contaminants. Astrophysical sources, such as Galactic dust emissions and synchrotron radiation, emit polarized signals that can significantly obscure the desired primordial $B$-mode polarization signal. Component separation between the cosmological signal and foreground emissions is a complex task that necessitates the implementation of sophisticated foreground cleaning techniques.
% }

In CMB analysis, there are various methods employed for component separation, broadly categorized into three groups: blind methods, parametric methods, and template removal methods. 
Blind methods, such as ILC (Internal Linear Combination)\cite{2003ApJS..148...97B, 2004ApJ...612..633E,2003PhRvD..68l3523T, 2009A&A...493..835D},  FastICA (Fast Independent Component Analysis)\cite{2002MNRAS.334...53M}, and CCA (Correlated Component Analysis)\cite{CCA-Bedini:2004dd}, make no assumptions about the foreground radiation. Parametric methods, like Commander\cite{2006ApJ...641..665E} and ForeGroundBuster\cite{fgb-Stompor:2008sf}, rely on assumptions about the components' models, such as their respective spectral laws, and fit the model parameters using the observation data. Template removal methods, such as SEVEM (Spectral Estimation Via Expectation Maximisation)\cite{2003MNRAS.345.1101M, 2008A&A...491..597L, 2012MNRAS.420.2162F}, constructing the foreground templates from the pairs of maps differences from the low and high frequency channels and obtaining the cleaned CMB maps by finding the coefficients to minimize the variance of map, SMICA (Spectral Matching Independent Component Analysis)\cite{2003MNRAS.346.1089D, 2008ISTSP...2..735C},  modelling the foregrounds as templates with spectral indices, power spectra and correlation between components. In most cases, these methods focus on quasi-fullsky. When it comes to small partial sky which is the common case for ground based experiment, extracting the right cleaned $B$-mode signal becomes more and more challenging because the resulting $B$ map (or power spectrum) can be strongly contaminated by $E$ to $B$ leakage \cite{2001PhRvD..65b3505L, 2003PhRvD..67b3501B}.

%\tred{Describe the problems or motivation: 1, For observations with high noise level, the general ILC methods will give rise to large bias due to ..., so we choose to use the cILC. 2, When compact sources occur in the observation maps, we need to mask out the strong compact sources very carefully, otherwise it will cause EB leakage problems, after we mask the compact source, we use the \textcolor{blue}{\sout{inpainting}} method to solve the EB leakage problem caused by holes.}

In our previous work\cite{2022JCAP...07..044Z}, we have shown that a Needlet space ILC working on the $EB$ leakage corrected $B$ map has a good performance. For a future ground-based experiment with a deep survey like CMB-S4\cite{S419}, the foreground residual can be well controlled which biases the tensor to scalar ratio ($r$) at the order of $10^{-3}$. However, the inherent limitation of the ILC algorithm itself is that it does not differentiate between instrument noise and foreground signals. As a result, in situations with low signal-to-noise ratio (SNR), the ILC methods tend to produce high foreground residual. So for $B$-mode analysis, the foreground residual finally bias the constraint of $r$. To improve the performance of ILC, i.e., reduce the foreground residual, additional information need to be involved to help ILC method more effective in distinguishing the foreground. 
One efficient approach is to incorporate an emission model of the foreground and enforce orthogonality of the ILC weights with respect to the foreground intensity. This can help in accurately distinguishing between the two components and improving the overall performance of the ILC method. That is the so called constrained ILC method, which was first proposed by the reference \cite{2011MNRAS.410.2481R}, and this method has also been widely employed to study CMB $B$-mode physics\cite{2021MNRAS.503.2478R, 2022JCAP...10..063G}.

%\tred{\sout{In this paper, we present an approach to address the challenges posed by low SNR in primordial $B$-mode measurements. We propose an enhanced version of needlet ILC (NILC) method on $B$ map, called constrained NILC, which incorporates additional prior foreground information.}} 
In this paper, we present an enhanced version of the NILC methodology for $B$-mode analysis. We utilize end-to-end simulations to validate the effectiveness of the improved method, demonstrating our approach with mock data and noise levels derived from WMAP\cite{2003ApJS..148....1B}, Planck\cite{2020A&A...641A...1P}, and a future ground-based experiment in the northern hemisphere.
These noise levels simulate the low SNR challenges encountered in actual observations in the near future.
Our results show that the level of foreground residual can be well controlled with constrained NILC. Comparing the performance of our constrained NILC method with the standard NILC method, we find a remarkable reduction in bias on $r$. Specifically, the bias is  reduced from $0.06$ to $0.007$, which is much smaller than the statistical error.

The paper is organized as follows. Section \ref{sec:metho} describes origin of the bias in ILC and introduces our proposed constrained NILC method. Section \ref{sec:simu} outlines the methodology employed for generating the mock data with given instrumental configuration. Section \ref{sec:implm} details how we implement our component separation method on the mock data to get the clean $B$ map and the subsequent constraint on $r$. Section \ref{sec:result} presents the results on map level, angular power spectral level as well as the $r$ constraint; section \ref{sec:conclusions} gives the conclusion.

\section{Motivation and Methodology}\label{sec:metho}
\subsection{The ILC estimator and its bias}\label{subsec:ICLbias}
The objective of the ILC method is to accurately isolate the specific emission of interest, such as CMB, from a set of multi-frequency maps. 
This technique assumes that the emission of interest remains consistent across different frequencies, allowing for the construction of a single cleaned map that predominantly represents the desired signal.

%The goal of ILC method is to construct the only one emission of interest (CMB) from the total multi-frequency maps, assuming the emission of interest independent of frequency. 
%In the classical ILC theoretical framework, it assumes that the observational frequency maps is a linear combination of the CMB signal, the foreground emission which scales with frequency, and the instrumental noise. 
%Because the blindness of the ILC methods, the contributions from non-CMB signal can be merged into one term. 
Suppose that all the frequency maps are calibrated to CMB units and the same resolution, the observation can be modeled as: 
\begin{align}
    \mathbf{y}(p)=\mathbf{1}s(p)+\mathbf{f}(p)+\mathbf{n}(p). \label{eq:obs}
\end{align}
Here, $\mathbf{y}(p)$, $\mathbf{1}$ and $\mathbf{n}(p)$ are $n_{\nu}$ dimensional vector, where $n_{\nu}$ denotes the number of frequency channels, $p$ can be any domain like pixel, harmonic, needlet, etc. The terms $\mathbf{f}$ and $\mathbf{n}$ describe the contribution from galactic diffuse foreground contamination and instrumental noise respectively. 
To minimize the impact of the ``junk'' term $\mathbf{f}$+$\mathbf{n}$, the ILC provides an estimator $\hat{s}_{\rm ILC}$ of $s$ by forming the linear combination $\hat{s}_{\rm ILC}(p) = \mathbf{w}^T \mathbf{y}(p)$,  which has minimum variance and also preserves the CMB signal with $\mathbf{w}^T\mathbf{1} = 1$. Such the weights are given by \cite{2003ApJS..148...97B}
\begin{align}
    \mathbf{w} = \frac{\hat{\mathbf{C}}^{-1}\mathbf{1}}{\mathbf{1}^T\hat{\mathbf{C}}^{-1}\mathbf{1}}, \label{eq:ilc_weight}
\end{align}
where $\hat{\mathbf{C}}$ is the empirical covariance matrix of the observations.

Suppose the number of frequency channel is larger than the combined number of foreground components and the CMB signal, and in an ideal scenario with no noise, the ILC estimator is unbiased. However, in the real world, bias always exists and the bias of ILC may come from different situations as follows:

\begin{itemize}
    \item The non-stationary of foreground and noise within the selected $p$ domain introduces a challenge as it leads to varying weights, which can introduce bias in the ILC method. To address this issue, various approaches have been explored to ensure the stationary of the field $\mathbf{f}$ and noise $\mathbf{n}$. One approach is to carefully choose appropriate $p$ domains by 1) decomposing the sky maps into several regions\cite{2003ApJS..148...97B}, 2) harmonic domain\cite{2003PhRvD..68l3523T}, 3) needlet domain\cite{2008MNRAS.383..539M}. Additionally, it is beneficial to scan the sky as uniformly as possible during observation.  
    \item The residual of ``junk'' term can be defined as $r=\mathbf{1}^T\hat{\mathbf{C}}^{-1}(\mathbf{f}+\mathbf{n}) / (\mathbf{1}^T\hat{\mathbf{C}}^{-1}\mathbf{1})$. When $\mathbf{f}$ and $\mathbf{n}$ are random field with the zero means, $\hat{s}_{\rm ILC}$ is unbiased, but when there exits non-zero higher order terms of auto-term and cross-term of $\mathbf{r}$, for example $\left<r^2\right> + 2\left<sr\right>$  will introduce bias\cite{2009A&A...493..835D}:
    
    a) The auto-term bias can the written as following with the assumption that there is no correlation between foreground and noise:  
    \begin{align}
        \left<r^2\right>=\frac{\mathbf{1}^T\hat{\mathbf{C}}^{-1}(\left<\mathbf{f}\mathbf{f}^T\right>+\left<\mathbf{n}\mathbf{n}^T\right>)\hat{\mathbf{C}}^{-1}\mathbf{1}}{(\mathbf{1}^T\hat{\mathbf{C}}^{-1}\mathbf{1})^2},
    \end{align}
    both the bias from foreground and noise increase with the noise level. However, we can easily remove noise bias by employing noise-only simulations with the knowledge of statistical property of noise which is described in Section \ref{sec:cmap2r}. Unfortunately, the same approach cannot be applied to the foreground. For low SNR $B$-mode physics, such bias is typically non-negligible. It is possible to obtain an unbiased ILC estimator by replacing the $\mathbf{C}$ with $\mathbf{C}-\mathbf{N}$, where $\mathbf{N}$ is the noise covariance, but this will leads to a substantial increase in the variance of CMB power\cite{2008A&A...487..775V}.

    b) The cross-term bias mainly arise from the difference between the true underlying covariance matrix $\mathbf{C}$ and the estimated empirical covariance matrix $\hat{\mathbf{C}}$. The bias values approximated amounts to $2(1-n_{\nu})\sigma^2_{s}/N_{p}$, which means that the ILC estimator underestimates the CMB power by roughly $2(1-n_{\nu})C_{\ell}/ N_{p} $, more details refer to \cite{2009A&A...493..835D}. 
    
\end{itemize}

In this paper, we focus on reducing the ILC bias with the case of low SNR. To minimize this bias, more additional information are needed to distinguish the Galactic foreground signal from the instrumental noise. In other words, extra constraints, which depend on the additional information, can be imposed on the weights. %\tblue{The bias induced by instrumental noise can be easily mitigated on the power spectrum by conducting multiple noise-only simulations using an appropriate noise model, allowing for debiasing. In the worst-case scenario, even without a noise model, it is still possible to divide the data into several segments based on time and calculate their cross spectra separately after performing map-making and ILC. In this situation, instrumental noise will not introduce bias, but residual foregrounds will remain.}

\subsection{Constrained Needlet ILC}
Suppose the emission law of all the Galactic foreground and the CMB signal are known, thus the observation can be modeled as 
\begin{align}
    \mathbf{y}(p) = \mathbf{As}(p) + \mathbf{n}(p).
\end{align}
Here we also assume that all maps have been re-smoothed to the same resolution. The mixing matrix $\mathbf{A}$ are known from the prior. $\mathbf{s}(p)$ is a column vector with $N_{\rm comp}$ elements denote the templates for all the components (CMB + foreground). And thus the term $\mathbf{n}(p)$ includes the instrumental noise only. The goal of constrained ILC is to reduce the impact of residual foreground signal while preserving the CMB signal. Thus the constraint becomes $\mathbf{A}^T\mathbf{w} = \mathbf{e}$ with $e_{\rm CMB} = 1$, $e_{\rm non-CMB} = 0$. The weights for minimizing the variance of the resulting map with the constraints can be solved using Lagrange multipliers\cite{2004ApJ...612..633E}. The solution is 
\begin{align}
    \mathbf{w} = \hat{\mathbf{C}}^{-1}\mathbf{A}(\mathbf{A}^T\hat{\mathbf{C}}^{-1}\mathbf{A})^{-1}\mathbf{e}. \label{eq:CILC_weights}
\end{align}

The foreground bias in this issue will arise bias of $\mathbf{A}$ in two main aspects: misunderstanding of the number of foreground components and the bias associated with the individual elements of $\mathbf{A}$. 

From our previous work, we have shown that the NILC on $B$ maps is very promising on $r$ constraint, so here we choose to estimate the weights in needlet domain. The following will describe the constrained NILC in detail, the multi-frequency data is in the form of needlet coefficients which given by
\begin{align}
    \boldsymbol{\beta}_{jk} = \sum_{\ell m} \mathbf{y}_{\ell m} h_{\ell}^{(j)}Y_{\ell m}\left(\hat{\xi}_{k}^{(j)}\right)
\end{align}
where $\mathbf{y}_{\ell m}$ is the spherical harmonic coefficients of the observational $\mathbf{y}(p)$ which can be arbitrary scalar or pseudo-scalar map. $\hat{\xi}^{(j)}_{k}$ is a set of grid points on sphere. In practice, they are exactly the center of the pixel $k$ in \texttt{Healpix} pixelization scheme with different $\texttt{NSIDE}(j)$. $h_{\ell}^{(j)}$ is the spectral windows satisfying $\sum_{j} (h_{\ell}^{(j)})^2 = 1$. The NILC estimation on needlet space is 
\begin{align}
    \hat{\beta}^{\rm NILC}_{jk} = \mathbf{w}^T_{jk}\boldsymbol{\beta}_{jk}. \label{eq:syn_beta}
\end{align}
The final estimation $\hat{s}_{\rm NILC}$ in real space is 
\begin{align}
    \hat{s}_{\rm NILC}(p) = \sum_{j}\sum_{\ell m}\hat{\beta}_{j, \ell m}^{\rm NILC} h_{\ell}^{(j)} Y_{\ell m}(p) \label{eq:NILC_estimator}
\end{align}
The weights in \eqref{eq:syn_beta} are calculated from the equation \eqref{eq:CILC_weights} with an empirical covariance $[\hat{\mathbf{C}}]_{jk}$
which are estimated by the formula
\begin{align}
    \hat{\mathbf{C}}_{jk} = \frac{1}{N_{\mathcal{D}}}\sum_{k'}\omega_{j}(k,k')\boldsymbol{\beta}_{jk'}\boldsymbol{\beta}_{jk'}^T. \label{eq:cov_estimation}
\end{align}
It means that the covariance matrix obtained by computing the mean of the needlet coefficients. In practice, the weights in \eqref{eq:cov_estimation} is a 2D Gaussian function. Due to the locality of the NILC method, it is convenient to extend equation \eqref{eq:CILC_weights} by using different mixing matrices $\mathbf{A}$ to compute weights at different positions. The estimation of $\mathbf{A}$ matrices from the multi-frequency maps will be shown in section \ref{sec:implm}.

% \subsection{Compact source masking and inpainting}

% \subsection{constrained ILC methods}

% \subsubsection{cILC in pixel domain on QU maps}

% \subsubsection{cILC in needlet domain on B maps}

\section{Map Simulations}\label{sec:simu}

\subsection{Sky model}

We adopted a sky model that is as close as possible to the real observation. It contains CMB signal, diffuse foreground emission composed of synchrotron, thermal dust, free-free and spinning dust. The polarized foreground maps are simulated by using the Planck Sky Model\cite{2013A&A...553A..96D} (PSM). Here are the details of the sky model:

\begin{enumerate}
    \item \textbf{CMB:} the input CMB maps are Gaussian realizations from a particular power spectrum obtained from the Boltzmann code \texttt{CLASS}\footnote{\url{http://lesgourg.github.io/class_public/class.html}}\cite{2011JCAP...07..034B}, using  Planck 2018 best-fit cosmological parameters \cite{2020A&A...641A...6P} with the tensor scalar ratio $r = 0$. The CMB dipole was not considered in the simulations.
    
    The effect of CMB distorted by weak gravitational lensing was also considered in our simulation. Weak lensing distorts a part of $E$-mode signal to $B$-mode and it is not a linear effect and therefore it is non-Gaussian. The precise way to simulate such lensing distortion effects is to rearrange the pixel on the map level. For this part, the package \texttt{LensPix}\footnote{\url{https://cosmologist.info/lenspix/}} is used to simulate it. the algorithm is presented in the reference\cite{2005PhRvD..71h3008L}, which distorts the primordial signal given a realization of lensing potential map from $C_{\ell}^{\phi\phi}$.
    
    %The existance of CMB lensing causes problems for primordial $B$-mode searching. 
    \item \textbf{Synchrotron:} Galactic synchrotron radiation is produced by charged relativistic cosmic rays that are accelerated by the Galactic magnetic field. At frequencies below 80 GHz, synchrotron radiation is the primary polarized foreground. In our simulations, we did not account for the effects of Faraday rotation, and instead modeled the synchrotron emission across frequencies using template maps based on specific emission laws. To simplify matters, we modeled the frequency dependence of synchrotron radiation in the CMB frequency range as a power-law function (in Rayleigh-Jeans units):
    \begin{align}
        \begin{bmatrix}
            T_{\rm sync}(\nu) \\ Q_{\rm sync}(\nu) \\ U_{\rm sync}(\nu)
        \end{bmatrix} \propto
        \begin{bmatrix}
            T_{template} \\ Q_{template} \\ U_{template}
        \end{bmatrix} \times \nu^{\beta_s}. \label{eq:sync_model}
    \end{align}
    The spectral index $\beta_s$ typically has a value of $-3$. When generating temperature maps, we utilized the 408 MHz radio continuum full-sky map of Haslam et al. \cite{1982A&AS...47....1H, 2015MNRAS.451.4311R} as our template. For polarization maps, we used the SMICA component separation of Planck PR3 (Planck Public Data Release 3) \cite{2020A&A...641A...4P} polarization maps as templates, with the same synchrotron spectral index.
    \item \textbf{Thermal dust:} At higher frequencies (above 80 GHz), the thermal emission from heated dust grains becomes the dominant foreground signal. In the PSM, we modeled the emission of thermal dust as modified black-bodies \cite{2014A&A...571A..11P},
    \begin{align}
        I_{\nu} \propto \nu^{\beta_{\rm d}}B_{\nu}(T_{\rm d}), \label{eq:dust_model}
    \end{align}
    where $T_{\rm d}$ is the temperature of the dust grains, $A_{\rm d}$ is the amplitude and $\beta_{\rm d}$ is the spectral index. $B_{\nu}(T_{\rm d})$ is the Planck black-body function at temperature $T_{\rm d}$ at frequency $\nu$. For polarization, the Stokes $Q$ and $U$ parameters (in ${\rm MJy}/{\rm sr}$ unit) are correlate with the intensity, 
    \begin{align}
        Q_{\nu}(\hat{\mathbf{n}})&=f(\hat{\mathbf{n}}) I_{\nu}\cos(2\gamma_d(\hat{\mathbf{n}})), & 
        U_{\nu}(\hat{\mathbf{n}})&=f(\hat{\mathbf{n}}) I_{\nu}\sin(2\gamma_d(\hat{\mathbf{n}})),
    \end{align}
    where $f(\hat{\mathbf{n}})$ and $\gamma_{\rm d}(\hat{\mathbf{n}})$ denote polarization fraction and polarization angle at different position $\hat{\mathbf{n}}$. We use the dust template in Planck PR3 at $353$ GHz in intensity and polarization as our template, which includes the template maps of the dust grains $T_{\rm  d}$, the amplitude for all $T,Q,U$ at 353 GHz, and the spectral indices.
    
    \item \textbf{free-free:} Free-free emission is produced by electron-ion scattering in interstellar plasma and is generally fainter than synchrotron or thermal dust emission. It is modeled as a power-law with a spectral index $\beta_{\rm f}$ of approximately $-2.14$ \cite{2016A&A...594A..10P} and is intrinsically unpolarized. Free-free emission only contributes to temperature measurements, and we used the free-free emission map from the \texttt{Commander} method \cite{2006ApJ...641..665E} in Planck 2015 data PR2 \cite{2016A&A...594A..10P} as our template.
    
    \item \textbf{Spinning dust:} Small rotating dust grains can produce microwave emission if they have a non-zero electric dipole moment, which can also be modeled as a power-law within the CMB observation frequency bands. We used the template spinning dust map from \texttt{Commander} in Planck PR2 as our input template. Spinning dust emission is weakly polarized, and in our simulation, the polarization fraction of spinning dust was set to 0.005.
    
    %\item \textbf{Point sources:} Radio sources are mainly modeled based on the surveys of radio sources of frequencies between $0.85$ GHz and $4.85$ GHz. The catalogues are detailed in the reference\cite{2013A&A...553A..96D}. The point sources are also modeled as a power law to approximate the spectra as $S\propto\nu^{-\alpha}$. The spectral indices $\alpha$ for each point sources are obtained from the large-scale surveys. The polarization of each source attributing to a polarisation degree is randomly drawn from the observed distributions for flat- and steep-spectrum sources at 20 GHz\cite{2004A&A...415..549R}, and a polarization angle is randomly drawn from a uniform distribution.
\end{enumerate}

\subsection{Instrumental configuration}

\begin{figure}[bt]
    \centering
    \subfloat[]{\includegraphics[width=0.45\textwidth]{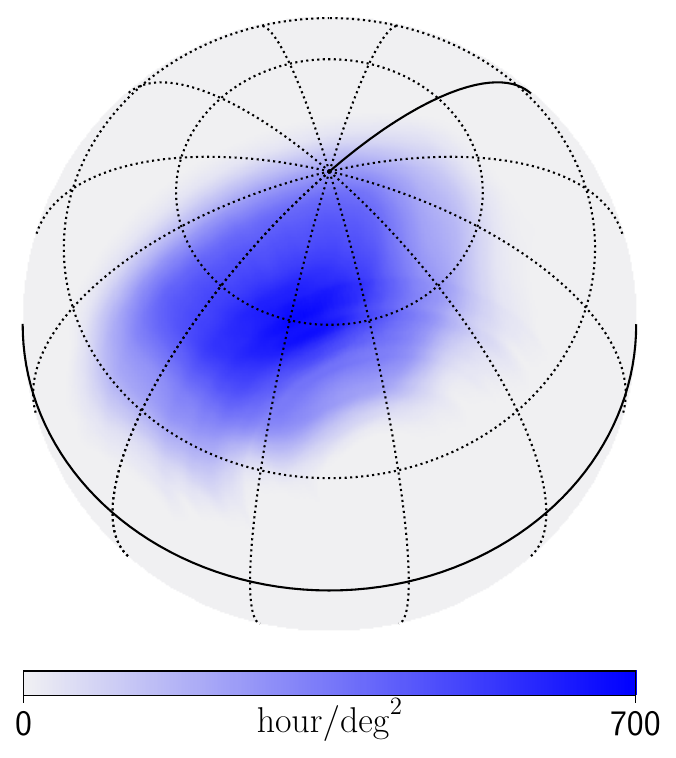}} \hskip 0.05\textwidth
    \subfloat[\label{subfig:skypatch}]{\includegraphics[width=0.45\textwidth]{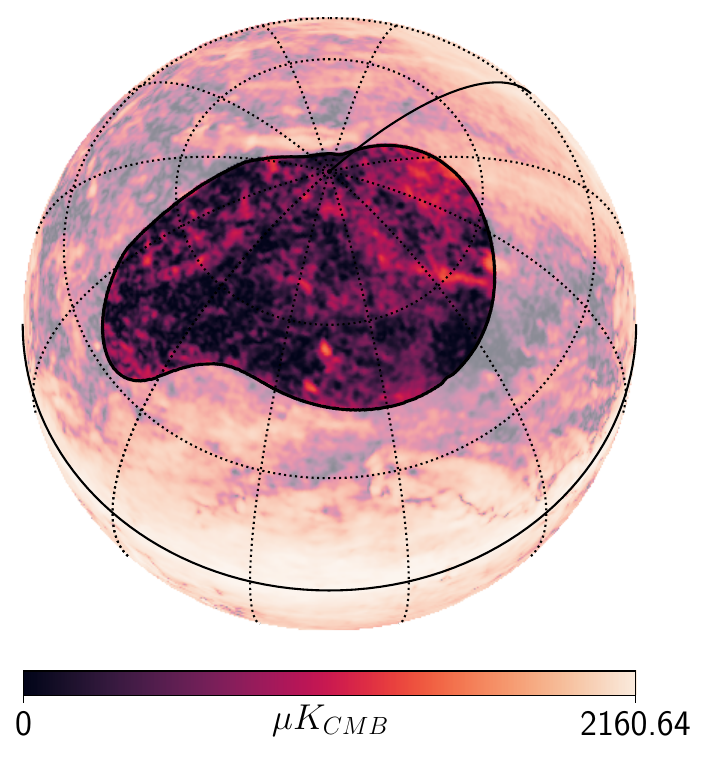}}
    \caption{Orthogonal projection of the hit-map and sky patches in Galactic coordinate. (a) shows the hit-map of the ground-based observation used in this work. (b) shows the sky patch we used for further analysis. The background of it is the polarization intensity $P=\sqrt{Q^2+U^2}$ map observed by Planck at $353$~GHz \cite{2020A&A...643A..42P} with $1^\circ$ resolution.}
    \label{fig:hitmap}
\end{figure}

In this work, we consider a ground-based observational experiment that can only cover a partial sky area. The scanning strategy is optimized for detecting primordial gravitational waves on Northern hemisphere, in other words, the scanning strategy was designed to perform a deep scan of a small sky region. The hit-map corresponding to the scanning strategy is shown in Figure \ref{fig:hitmap}. The deep scanning sky patch is centered at RA~$10$h$04$m, Dec.~$61.49^\circ$. The effective area of the sky-patch is about $7.5\%$ of the full-sky.

Considering the atmospheric window, we have set up two observation frequency bands of 95 and 150~GHz for the ground-based experiments. In order to achieve better foreground removal results for ILC, we also simulated some space-based observed sky maps and taking them as the inputs. The maps including the K and V bands of WMAP, as well as all bands of Planck High Frequency Instrument (HFI). The beam and noise information of them was obtained from the LAMBDA website\footnote{\url{https://lambda.gsfc.nasa.gov}}. 

%Two configurations are adopted in this work, and both of them are assumed as future ground-based CMB experiments. For primordial gravitational waves detection, it should carry out a deep survey on a clean sky patch and accumulate enough scanning, and along this guidance we select two clean patches, which are distributed in northern hemisphere and southern hemisphere respectively. For the northern sky patch, we choose the area centered at ($RA\ 13.13h$, $Dec\ 55^\circ$), and it is about  $7\%$ sky coverage where it appears to be the patch with lowest galactic foreground according to the Planck 353 GHz polarization intensity map. For the southern sky patch, a circular patch was choosed with roughly $3\%$ sky coverage which is adopted in CMB-S4 primordial $B$ mode forecasting. The sky patches with Planck 353 GHz polarization emission intensity map is shown in Figure \ref{fig:skypatch_mask}\subref{subfig:skypatch} and \ref{fig:skypatch_mask}\subref{subfig:skypatchCMBS4}. 

\begin{table}
    \centering
    \begin{threeparttable}
        \caption{Instrumental configuration} \label{tab:inst_config}
        \begin{tabular}{cccc}
        \toprule
            Frequency bands & Beam size & Polarization map-depth\tnote{*} & Type \\
            (GHz) & (arcmin) & ($\mu{\rm K}\cdot{\rm arcmin}$) & \\
        \midrule
            23 & 52.8 & $271$ & \multirow{2}*{WMAP}\\
            61 & 21.0 & $141$ \\ \midrule[.2pt]
            95 & 19.0 & $20$  & \multirow{2}*{Ground-based}\\
            150 & 11.0 & $31$ \\\midrule[.2pt]
            100 & 9.94 & $105$ & \multirow{4}*{Planck HFI}\\ 
            143 & 7.04 & $63$ \\
            217 & 4.66 & $95$ \\
            353 & 4.41 & $394$ \\
        \bottomrule
        \end{tabular}
        \begin{tablenotes}
            \item [*] The average value within the sky-patch
        \end{tablenotes}
    \end{threeparttable}
\end{table}

The frequency bands and the related beam resolution are listed in Table \ref{tab:inst_config}. All the maps are in the \texttt{HealPix} pixelization scheme at $\texttt{NSIDE}=512$. The noise in the sky patch is not homogeneous. For simulated sky maps of ground-based observations, the noise variance on each pixel depends on the hit-map, which shown in Figure \ref{fig:hitmap}. The variance is given by 
\begin{align}
    \sigma^2_p = \frac{{\rm NET}^2}{H_p\Omega_p},
\end{align}
where $H_p$ denotes the value of hit-map at pixel $p$ (in ${\rm hour}/{\rm deg^2}$ unit) and $\Omega_p$ is the solid angle of pixel $p$. 
In the case of simulating space-based observed sky maps, the noise variance map can be obtained from the LAMBDA website.

% The noise distribution is assumed to be homogeneous among the patch. The map depth for each frequency band is also listed in the same table.
% The final observed multi-frequency maps were generated by adding the noise map realized from map depth and the simulated sky maps at each frequency. 
% All the maps are pixelized in HealPix format at $\texttt{NSIDE}=512$. We also use the same map depth to generate 500 noise-only simulations which can be used to debias the noise at angular power spectral level. 

\section{Foreground removal implementation and r constraint}\label{sec:implm}
\subsection{Foreground removal implementation}

To perform a constrained NILC on the simulated observational sky maps, 
there is still a key issue to be solved, 
which is how to obtain the mixing matrix $\mathbf{A}$ in equation \eqref{eq:CILC_weights}. 
Generally speaking, the term $\mathbf{A}$ encodes the emission law of the foreground. 
Based on the current understanding of the foreground model, 
only synchrotron radiation and thermal dust emission can contribute polarization signal. 
From the synchrotron and dust model equations \eqref{eq:sync_model} and \eqref{eq:dust_model}, 
it appears that we need to know these three parameters $\beta_{\rm s}, \beta_{\rm d}$ and $T_{\rm d}$ in advance. 

In our implementation, the parameter $T_{\rm d}$ is fixed to be $19.6$~K according to the results of Planck\cite{Planck:2013ltf}. 
As for $\beta_{\rm s}$ and $\beta_{\rm d}$, 
We will fit them from the observational intensity and polarization sky maps.
The fitting is based on three assumptions:
1. The foreground emission is dominated by large-scale signals.
2. The spectral indices of any foreground components modeled by power-law change very little on small scales.
3. The frequency dependence of foreground emission is the same for both intensity and polarization signals.
We fit the values of these two parameters at each pixel using the $8\ (frequency\ channel) \times3\ (T, Q, U)$ observed sky map. The model at each pixel $p$ is 
\begin{align}
    T_{\nu}(p) &= A_{\rm s}(p)\frac{g_{\rm RJ}(\nu)}{g_{\rm RJ}(\nu_{\rm s})}\left(\frac{\nu}{\nu_{\rm s}}\right)^{\beta_{\rm s}} + A_{\rm f}(p)\frac{g_{\rm RJ}(\nu)}{g_{\rm RJ}(\nu_{\rm s})}\left(\frac{\nu}{\nu_{\rm s}}\right)^{-2.14} + A_{\rm sd}(p)\frac{g_{\rm Jy}(\nu)}{g_{\rm Jy}(\nu_{\rm s})}\frac{f_{\rm sd}(\nu)}{f_{\rm sd}(\nu_{\rm s})}\nonumber\\
    & + A_{\rm d}(p)\frac{g_{\rm Jy}(\nu)}{g_{\rm Jy}(\nu_{\rm d})}\frac{B_{\nu}(T_{\rm d})}{B_{\nu_{\rm d}}(T_{\rm d})}\left(\frac{\nu}{\nu_{\rm d}}\right)^{\beta_{\rm d}(p)} + T_{\rm CMB}(p),\label{eq:modelT}\\
    Q_{\nu}(p) &= Q_{\rm s}(p)\frac{g_{\rm RJ}(\nu)}{g_{\rm RJ}(\nu_{\rm s})}\left(\frac{\nu}{\nu_{\rm s}}\right)^{\beta_{\rm s}} + Q_{\rm d}(p)\frac{g_{\rm Jy}(\nu)}{g_{\rm Jy}(\nu_{\rm d})}\frac{B(\nu, T_{\rm d})}{B(\nu_{\rm d}, T_{\rm d})}\left(\frac{\nu}{\nu_{\rm d}}\right)^{\beta_{\rm d}(p)} + Q_{\rm CMB}(p),\label{eq:modelQ}\\
    U_{\nu}(p) &= U_{\rm s}(p)\frac{g_{\rm RJ}(\nu)}{g_{\rm RJ}(\nu_{\rm s})}\left(\frac{\nu}{\nu_{\rm s}}\right)^{\beta_{\rm s}} + U_{\rm d}(p)\frac{g_{\rm Jy}(\nu)}{g_{\rm Jy}(\nu_{\rm d})}\frac{B(\nu, T_{\rm d})}{B(\nu_{\rm d}, T_{\rm d})}\left(\frac{\nu}{\nu_{\rm d}}\right)^{\beta_{\rm d}(p)} + U_{\rm CMB}(p).\label{eq:modelU}
\end{align}
Here $\nu_{\rm s} \equiv 23$~GHz and $\nu_{\rm d} \equiv 353$~GHz. The factors $g_{\rm RJ}(\nu)$ and $g_{\rm Jy}(\nu)$ denote the conversion factors from antenna temperature $\mu \rm{K}_{\rm RJ}$ or flux density Jy/sr to CMB unit $\mu {\rm K}_{\rm CMB}$ at frequency $\nu$. 
We considered the spinning dust emission are unpolarized in our fitting model. $f_{sd}$ is the external template calculated from \texttt{SpDust2} \cite{2009MNRAS.395.1055A}.
Totally there are $12$ free parameters in the model include the amplitudes and the spectral indices of the foreground, and the CMB signal. 

In order to achieve better fitting results, all the observational maps are re-smoothed to the same resolution $52.8$~arcmin and down graded the \texttt{NSIDE} to $8$. This step can average out the noise and CMB signal on each pixel, thereby improving the fitting results. 
After smoothing and downgrading, the spectral index for each pixel is the average of the true spectral index values within the pixel area. It can be easily figure out by perform Taylor expansion
\begin{equation}
    s_{\rm FG}(p) = \int {\rm d}\bm{n} A(\bm{n}) \nu^{\beta(\bm{n})} \simeq \bar{A}\nu^{\bar{\beta}} + \bar{A}(\ln\nu)^2\int(\beta(\bm{n}) - \bar{\beta})^2 + \dots,
\end{equation}
with $\bar{A} \equiv \int {\rm d}\bm{n} W(\bm{n}) A(\bm{n})$ and $\bar{\beta} \equiv \int {\rm d}\bm{n} W(\bm{n}) \beta(\bm{n})$, where $W(\bm{n})$ is the weight that encodes the information for smoothing and downgrading. Based on our aforementioned assumptions, we only consider the zeroth-order term.

\begin{figure}[tb]
    \centering 
    \subfloat[]{\includegraphics[width=0.45\textwidth]{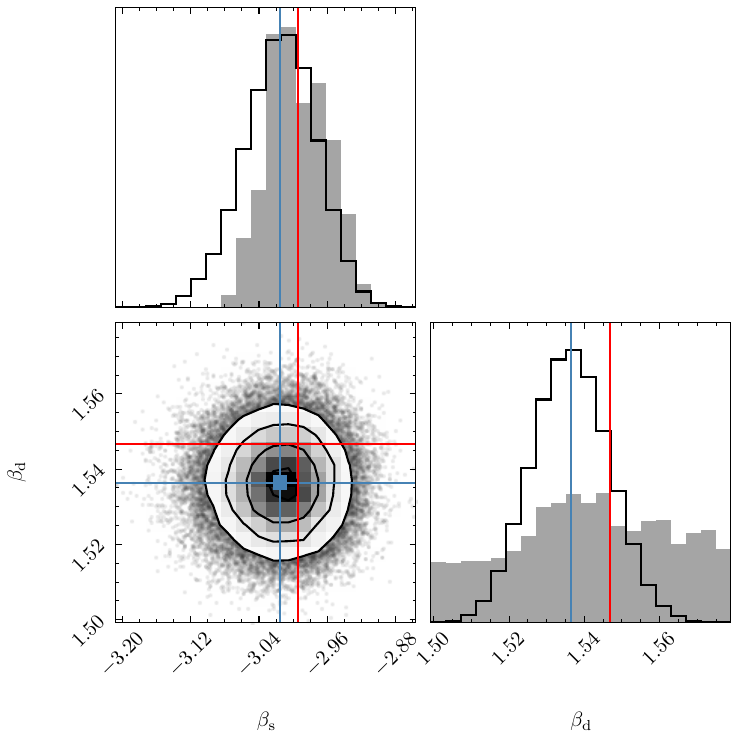}}
    \subfloat[]{\includegraphics[width=0.45\textwidth]{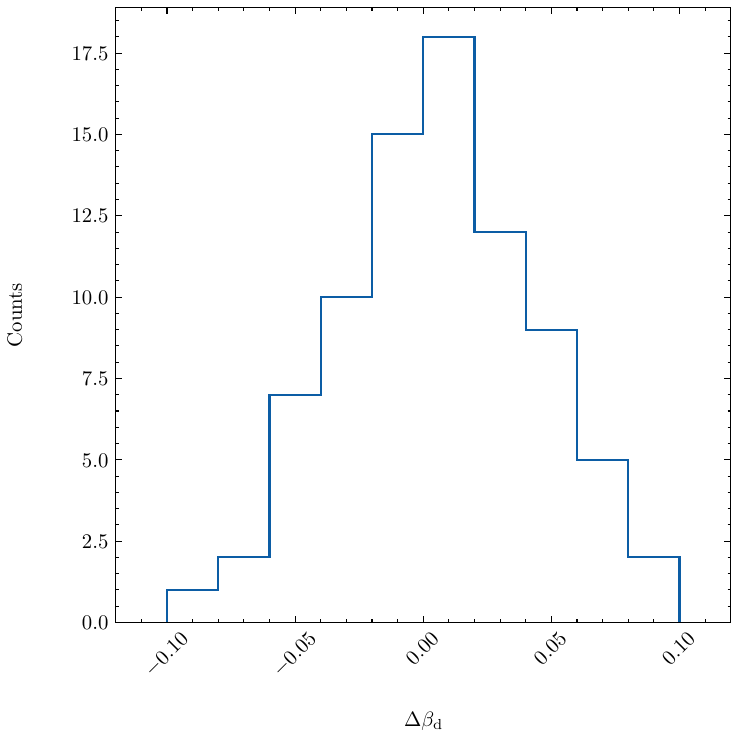}}
    %\caption{(a) The posterior distribution for the spectral index of synchrotron and thermal dust shown in Equation (\ref{eq:modelT} -- \ref{eq:modelU}), as obtained through the MCMC fitting algorithm for $8\times 3$ map-level data points on one pixel. The fitting results are $\beta_{\rm s} = -3.01^{+0.10}_{-0.09}$ and $\beta_{\rm d} = 1.536^{+0.011}_{-0.011}$ within $95\%$ CL. (b) The central values of parameter $\beta_{\rm d}$ as a map in the resolution $\texttt{NSIDE} = 8$.}
    \caption{(a) The solid step-line histogram and the contour represent the posterior distribution for the spectral index of synchrotron and thermal dust, as shown in Equations \eqref{eq:modelT} - \eqref{eq:modelU}. These distributions were obtained through the MCMC fitting algorithm for $8\times 3$ map-level data points within one pixel. The histogram, filled with gray color, shows the actual spectral indices in the fitting area, while the red line represents the averaged spectral indices in the area. The fitting results are $\beta_{\rm s} = -3.01^{+0.10}_{-0.09}$ and $\beta_{\rm d} = 1.536^{+0.011}_{-0.011}$ within $95\%$ CL. The central value is represented by the blue line in the plot. (b) Shows the difference in $\beta_{\rm d}$ between the blue and red lines as depicted in (a). A total of 81 pixels at $N_{\text{side}} = 8$ were fitted.}
    \label{fig:post_beta}
\end{figure}

The likelihood for each pixel, denoted as $\mathcal{L}(p)$, is given by
\begin{align}
    -2\log \mathcal{L}(p) = \sum_{\nu} \frac{\left[\hat{T}_{\nu}(p) - T_{\nu}(p)\right]^2}{\sigma^2_T(\nu, p)} + \frac{\left[\hat{Q}_{\nu}(p) - Q_{\nu}(p)\right]^2}{\sigma^2_P(\nu, p)} + \frac{\left[\hat{U}_{\nu}(p) - U_{\nu}(p)\right]^2}{\sigma^2_P(\nu, p)} + const. \label{eq:loglike_map}
\end{align}
Here $\sigma^2_T$ and $\sigma^2_P$ represent the variances of instrumental noise of the temperature map and polarization map, respectively. The variances are recalculated at $\texttt{NSIDE} = 8$. For each pixel, we sampled the likelihood \eqref{eq:loglike_map} by using Markov Chain Monte Carlo (MCMC). For efficiency purpose, we use the python package \texttt{NumPyro}\footnote{\url{https://github.com/pyro-ppl/numpyro}}\cite{phan2019composable, bingham2019pyro} to sample the likelihood. When running MCMC, the priors for all parameters are uniform distributed. For $\beta_{\rm s}$ and $\beta_{\rm d}$, we use the ranges $\beta_{\rm s} \in [-4,-2]$ and $\beta_{\rm d} \in [1, 3]$. For every pixel $p$, we take the best-fit value sampled from MCMC as the fitting results $\beta_{\rm s}(p)$ and $\beta_{\rm d}(p)$. Thus, $\beta_{\rm s}$ and $\beta_{\rm d}$ are the maps with the resolution $\texttt{NSIDE}=8$ after the fitting. An fitting example is given in Figure \ref{fig:post_beta} shows the fitting results of some of the parameters. 

According to the current understanding of the polarization foregrounds\cite{2016A&A...594A..10P, 2020A&A...641A..11P}, the model for polarized signal includes CMB, synchrotron and thermal dust emission. 
This suggests that we can apply two other separate constraints on the NILC method. 
One for removing synchrotron emission signal $\sum_{\nu}w_{\nu}A_{\nu, {\rm s}} = 0$ where $A_{\nu, {\rm s}}$, which is the element of mixing matrix $\mathbf{A}$ and given by Equation \eqref{eq:sync_model} and another for removing thermal dust emission signal $\sum_{\nu}w_{\nu}A_{\nu, {\rm d}} = 0$ where $A_{\nu, {\rm d}}$ is given by Equation \eqref{eq:dust_model}. 
It should be noted that the conversion factors to CMB unit are needed when calculating $A_{\nu, {\rm s}}$ and $A_{\nu, {\rm d}}$. 
Due to the scarcity of low-frequency observation data, fitting the model (\ref{eq:modelT} -- \ref{eq:modelU}) independently on each pixel results in a substantial error in the estimation of parameter $\beta_{\rm s}$ for all pixels.
In this case, the 95\%~CL for $\beta_{\rm s}$ is about $0.5$.  
Therefore, in order to mitigate the error, we impose the constraint that parameter $\beta_{\rm s}$ remains constant across all pixels during the fitting process. 
However, it is important to note that parameter $\beta_{\rm d}$ is allowed to vary across different pixels, as it represents distinct parameters for each pixel.
In this case, the signal-to-noise ratio for the parameter $\beta_{\rm s}$ is improved, reaching a value of $0.1$ at a 95\% CL.
% take $\beta_{\rm s}(p)$ as a nuisance parameter and use only $\beta_{\rm d}(p)$ to constraint dust emission for every pixel. For synchrotron emission, in order to improve the signal-to-noise ratio, the spatial information has been discarded. In other words, only taking the average value of each sky map as input data and then performing the MCMC sampling. Each pixel share the same value $\beta_{\rm s}$ obtained from the MCMC sampling chains. 
The posterior distribution for $\beta_{\rm s}$ and $\beta_{\rm d}$ are shown in Figure \ref{fig:post_beta}.

To better recover the $B$-mode CMB signal, the constrained weights $\mathbf{w}_{jk}$ in Equation \eqref{eq:CILC_weights} are calculated on $B$ maps defined as 
\begin{align}
 B(p) = \sum_{\ell m} a_{\ell m}^B Y_{\ell m}(p).   
\end{align}
The $EB$ leakage\cite{2001PhRvD..65b3505L} due to limited sky coverage is also corrected by a so-called template subtraction method\cite{PhysRevD.100.023538} when obtain $B$ maps from observational $QU$ maps. 
We have provided a detailed description of the steps for the EB leakage correction in the appendix.
The correction method has been proven to be the optimal blind correction on the pixel domain in the literature\cite{2019JCAP...04..046L}. 
We have outlined the specific correction steps and results in Appendix~\ref{sec:appendix}.
We show that the ILC methods performs slightly better on such leakage-free $B$-family maps\cite{2022JCAP...07..044Z}.  In addition, because of the localization of NILC, the mixing matrix $\mathbf{A}$ can vary with spatial position as $\mathbf{A}_{jk}$. In order to match the \texttt{NSIDE} of $\beta_{\rm d}(p)$ maps with $\texttt{NSIDE}(j)$ for every needlet band $j$, we use \texttt{map2alm} and \texttt{alm2map} routines in \texttt{healpy}\footnote{\url{https://github.com/healpy/healpy}}\cite{2005ApJ...622..759G, Zonca2019} package to up-grade the resolution.

\begin{figure}[bt]
    \centering
    \includegraphics{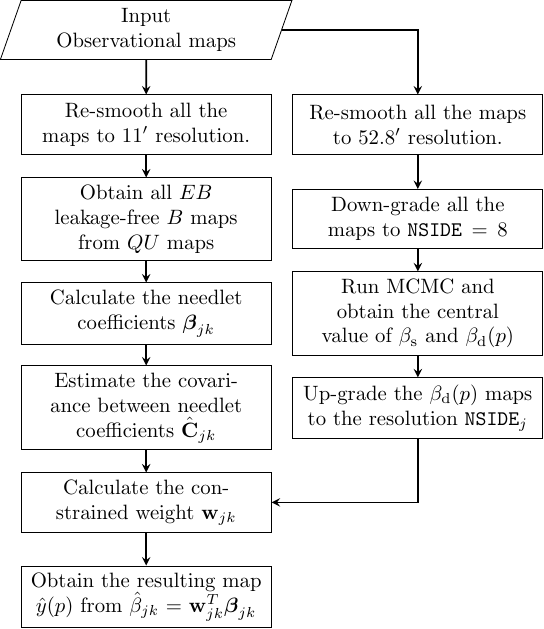}
    \caption{The flow chart of the whole constrained NILC process. }
    \label{fig:flowchart}
\end{figure}

In order to more intuitively describe how the constrained NILC process is executed from beginning to end, Figure \ref{fig:flowchart} shows the flow chart of the entire data processing process.

\subsection{From cleaned map to r constraint}\label{sec:cmap2r}
After obtaining the clean $B$ map through the constrained NILC method, the post-mapmaking analysis procedures are quite straightforward. In general, the subsequent processing involves the following steps.

\textbf{Angular power spectra estimation:} The pseudo-power spectra are estimated by using the python package of NaMaster\footnote{\url{https://github.com/LSSTDESC/NaMaster}}\cite{2019MNRAS.484.4127A}. In order to avoid instability caused by signal jumps at the edges of the sky regions, a mask that was slightly smaller than the mask shown in Figure \ref{fig:hitmap}\subref{subfig:skypatch} by about $2$ degrees is used. The power spectra are binned every $30$ multipole moments from $\ell=35$ to $215$. 

\textbf{Noise debias:} We use the $50$ noise only simulations to go through the same constrained NILC processing while keeping the weights unchanged to reconstruct $50$ reconstructed noise only maps, then estimate the mean value the band-powers $N_{\ell}$ over these noise simulations, finally subtract it from the band-power calculated from the resulting map of ILC. After subtraction, the band-power is considered to be noise-unbiased. Also, the $50$ noise only simulation band-powers are used to estimate the noise covariance matrix.

\textbf{Constraint on $r$:} To show how good our method is, we also use the calculated unbiased clean spectra to constrain the tensor to scalar ratio $r$. The full posterior for the individual band powers is non-Gaussian. However, for high enough multipoles (usually $\ell\ge30$) the central limit theorem justifies a Gaussian approximation\cite{2008PhRvD..77j3013H}. The lowest multipole $\ell$ in our binning scheme is $35$ thus the Gaussian approximation is valid:
\begin{align}\label{eq:likelihood}
    -2\ln\mathcal{L} = (\hat{C}_{\ell_b}^{BB}-C_{\ell_b}^{BB})^T{\rm Cov}(\hat{C}_{\ell_b}^{BB},\hat{C}_{\ell_b}^{BB})^{-1}(\hat{C}_{\ell_b}^{BB}-C_{\ell_b}^{BB}) +  const.
\end{align}
where the observed power spectrum is denoted as $\hat{C}_{\ell_b}^{BB}$. $C_{\ell_b}^{BB}$ denotes the theoretical power spectrum which was calculated by \texttt{CAMB} code. The likelihood only concerns the $BB$ spectra at $\ell\ge30$, where the primordial gravitational wave and lensing effect matters, so we fixed all the parameters to the best fit value from Planck-2018\cite{2018arXiv180706209P}, while only free $r$ with prior $r \in [0, 1]$. The covariance term is estimated from the noise only simulations. We sample this likelihood for these two parameters using the \texttt{CosmoMC}\cite{2002PhRvD..66j3511L}\footnote{\url{https://cosmologist.info/cosmomc/}} package. The posteriors are summarized by \texttt{GetDist}\cite{Lewis:2019xzd}\footnote{\url{https://pypi.org/project/GetDist/}} package.

\section{Results and discussion}\label{sec:result}
In this section, we present the numerical results obtained from the constrained NILC analysis. To demonstrate the stability of this method, we performed $500$ independent simulations. We construct the cleaned CMB maps estimated by constrained NILC foreground removal, as well as the residual foreground on the resulting map. We compute the angular power spectra of each of these maps and compare them to see its efficiency. As a comparison, we also perform the standard NILC method without any extra constraint on the same simulated data. As expected, the bias of standard NILC is larger than that of constrained NILC. 

% We discuss the influence of point sources on ILC analysis methods. Finally, we discuss the foreground residual from ILC methods and its influence on uncertainty on $r$.

% maps of different frequency bands, which is the difference map between the estimated foreground maps and the simulated input foreground maps.

\subsection{CMB signal reconstruction and foreground residual}

\begin{figure}[t]
    \centering
    \subfloat[]{\includegraphics[width=0.24\textwidth]{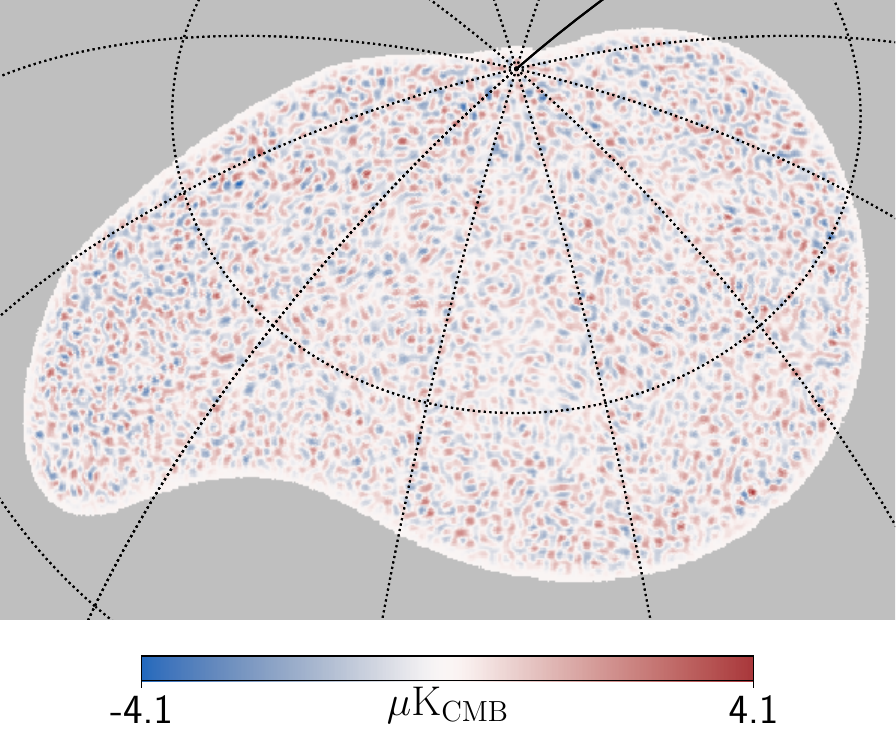}}
    \subfloat[]{\includegraphics[width=0.24\textwidth]{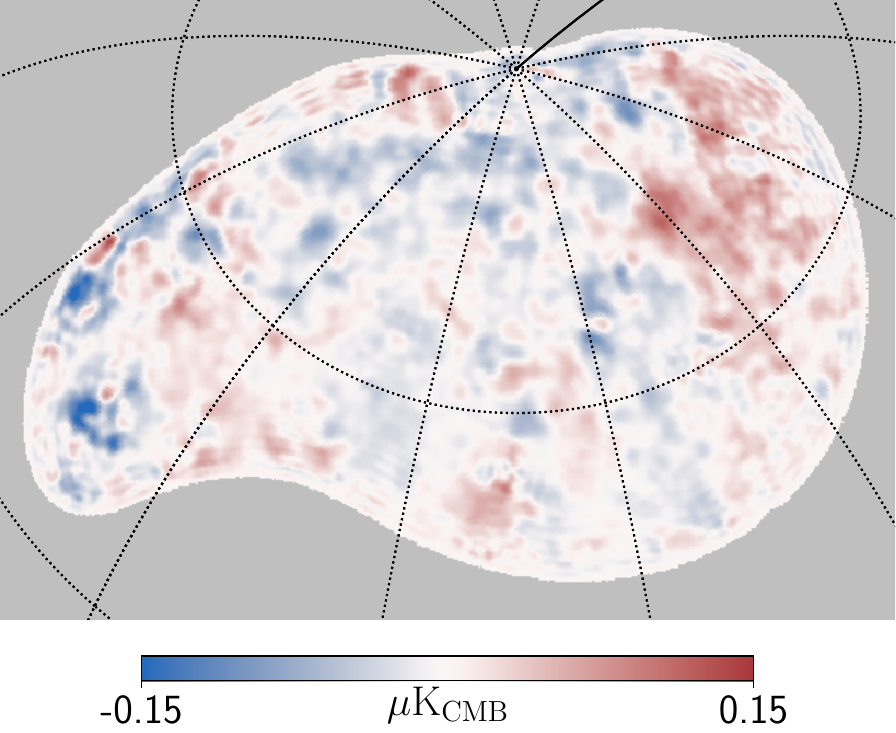}} \hskip 0.01\textwidth
    \subfloat[]{\includegraphics[width=0.24\textwidth]{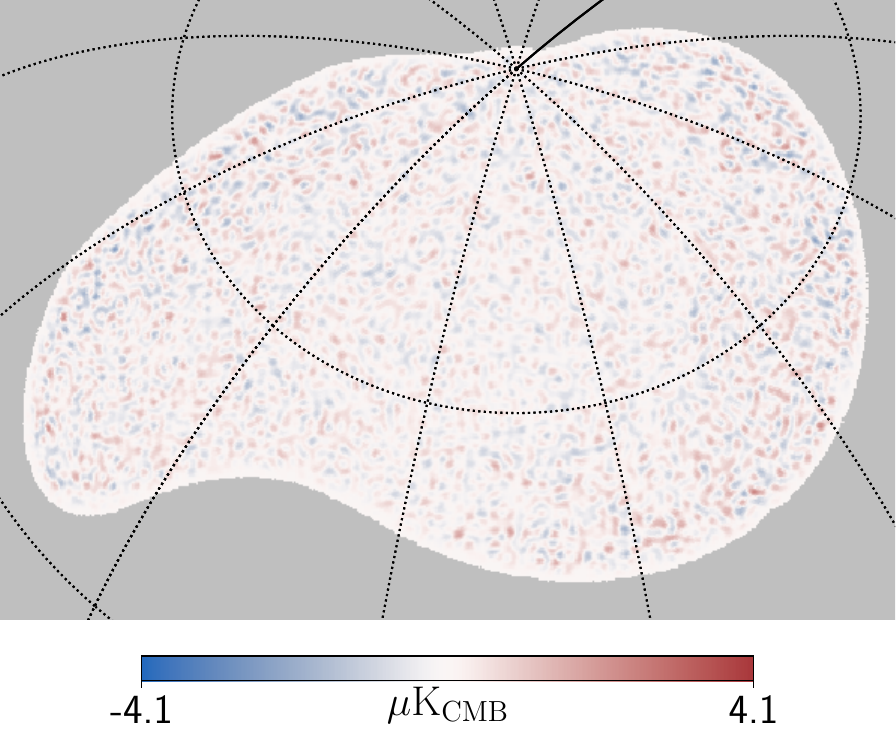}}
    \subfloat[]{\includegraphics[width=0.24\textwidth]{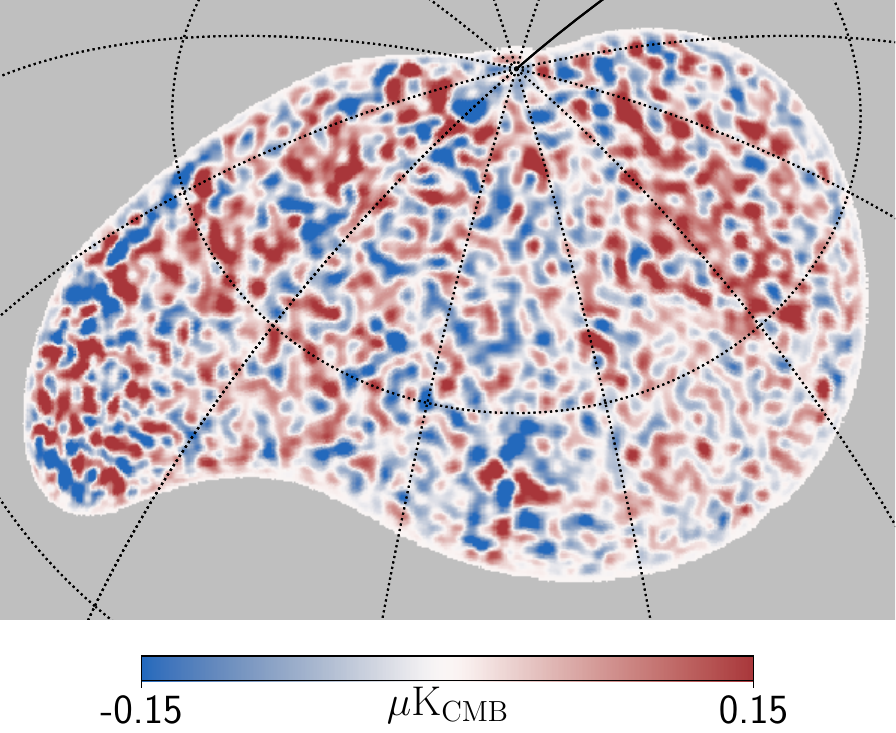}} \hskip 0.01\textwidth
    \\
    \subfloat[]{\centering\includegraphics[width=0.49\textwidth, origin=c]{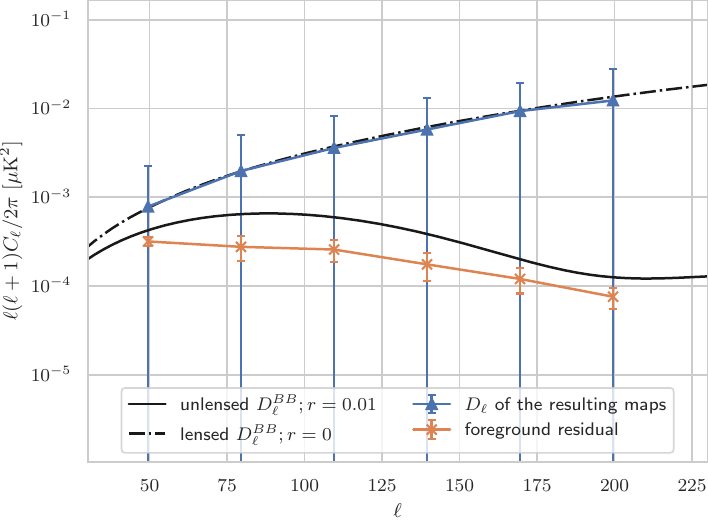}}
    \subfloat[]{\centering\includegraphics[width=0.49\textwidth, origin=c]{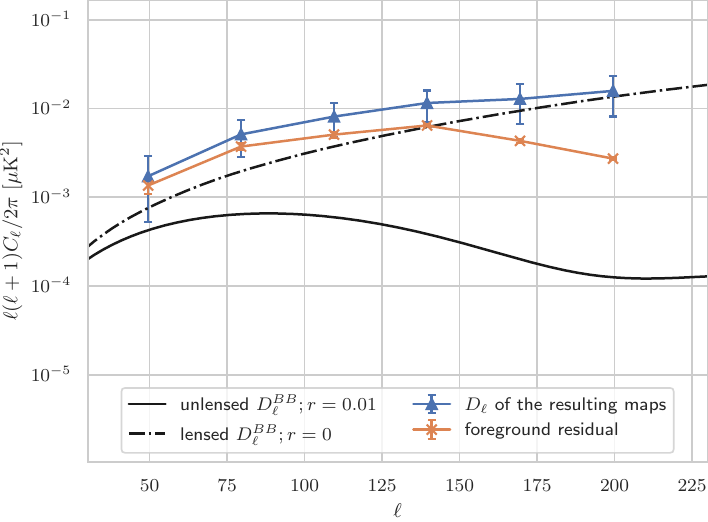}}

    \caption{The resulting and residual foreground $B$-mode maps obtained by constrained NILC and standard NILC as well as the band-powers of them. All the maps are smoothed with a Gaussian beam of $11$~arcmin, and the edges of them are apodized by $2$~degrees. (a) and (b) show the resulting and residual foreground maps of constrained NILC. The band-powers of these two maps are shown in (e). (c), (d) and (f) show the same results but for standard NILC. The dot-dashed line is the input lensed $B$-mode power spectrum with $r=0$ and the solid curve is an unlensed $B$-mode power spectrum with $r=0.01$ for comparison. All the band-powers in (e) and (f) are the averaged band-powers from $500$ independent simulations. The error-bars show the standard deviation of the simulations.}
    \label{fig:CNILC_result}
\end{figure}

The foreground residual maps can be obtained by applying the same weights obtained from constrained NILC or standard NILC to the input foreground maps. 
The resulting and foreground residual $B$-mode maps are shown in Figure \ref{fig:CNILC_result}. 
When comparing the results of the constrained NILC and standard NILC methods at the map level, it's easy to find that qualitatively, the resulting map of the constrained NILC exhibits slightly larger fluctuations compared to the standard NILC. 
However, this trend is reversed when considering the foreground residuals. More quantifiable results can be observed at the power spectrum level. 
It's clear that the foreground residual of the constrained NILC method is much smaller. From the power spectra analysis, the bias caused by foreground residual in the constrained NILC method is estimated to be around $r\sim 10^{-2}$. On the other hand, the standard NILC method exhibits a significantly larger foreground residual, estimated to be approximately at the order of $r=10^{-1}$.

One notable observation from the residual foreground maps is that the residual foreground of standard NILC exhibits a higher degree of small-scale structure compared to the residual of constrained NILC. This indicates that the foreground and the instrumental noise are indistinguishable especially on small scale where the noise dominate. The root of the problem lies in the fact that standard NILC lacks any prior information regarding noise and foreground components , as mentioned in Section \ref{subsec:ICLbias}. Adding prior information helps to get results with lower foreground residuals. But if the foreground and noise are considered as a whole ``junk'' term, then the overall residual will become larger. This is why the fluctuations in the resulting map of constrained NILC are greater than that of standard NILC. In a word, adding the extra constraints on NILC can help reduce the bias, but slightly increases uncertainty. 

\subsection{The constraint on r}

\begin{figure}[bthp]
    \centering
    \includegraphics[width=0.45\textwidth]{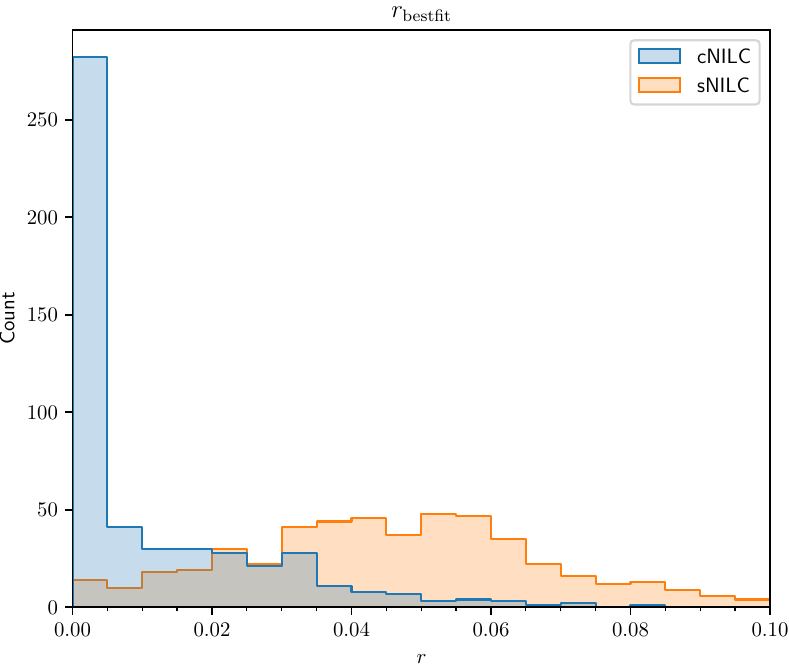}
    \includegraphics[width=0.45\textwidth]{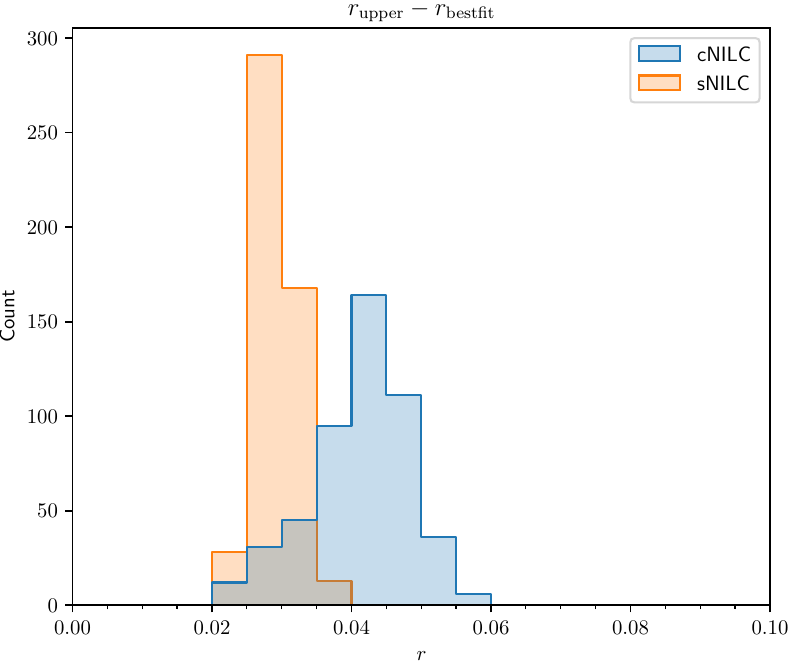}
    \caption{The left shows the distribution of the best-fit values $r_{\rm bestfit}$ of tensor-scalar ratio $r$. The right shows the distribution of the difference between one-sided upper bounds $r_{\rm upper}$ and the best-fit values $r_{\rm bestfit}$. Both histogram are sampled from $500$ independent simulations.}
    \label{fig:rlimits}
\end{figure}

One of the most important scientific goal of observing CMB $B$-mode polarization is to find out the polarized signal produced by the primordial gravitational waves. 
Figure \ref{fig:rlimits} illustrates the distribution of the best-fit values (left) and the differences between the $95\%$ upper limits and the best-fit values (right) for both the standard NILC and constrained NILC methods, based on $500$ independent simulations. 
It is evident that the constrained NILC method has a significantly smaller bias, less than $5\times10^{-3}$, compared to the standard NILC method, which exhibits a bias of approximately $r\sim 0.05$. 
Additionally, as anticipated, the figure also demonstrates that the difference, which represents the level of uncertainty, increases accordingly.
Indeed, for the standard NILC method, the $2\sigma$ uncertainty is approximately $0.025$, while for the constrained NILC method, it increases to around $0.045$. This discrepancy arises because the weights used in the constrained NILC method are optimized specifically for minimizing the foreground residual, rather than minimizing the overall impact of foregrounds and noise. As a result, there is a trade-off between reducing bias and increasing uncertainty in practice.
In conclusion, the obtained fitting results align with the analysis of residual foregrounds conducted in the previous section. 
% \includegraphics[width=0.5\textwidth]{Figures/bestfit_r_dist_CNILC.pdf}
% \includegraphics[width=0.5\textwidth]{Figures/bestfit_r_dist_NILC.pdf}
% \includegraphics[width=0.5\textwidth]{Figures/rlimits/bestfit_r_dist_CNILC.pdf}
% \textcolor{red}{To show the improvement of NcILC on B Maps method, I think we need compare this method with NILC on B maps from previous paper and the PcILC on QU maps.
% }

\section{Conclusions}\label{sec:conclusions}
The effectiveness of the ILC method for foreground removal at low SNR is limited by its inherent constraints. However, by incorporating additional prior information and transforming it into a semi-blind method, the residual foreground can be significantly reduced. 
In this work, we propose an improved needlet ILC foreground removal method, referred to as constrained NILC, which specifically designed for ground-based CMB observations searching for $B$-mode. 
The complete workflow of the constrained NILC method is depicted in Figure \ref{fig:flowchart}. 

To validate the effectiveness of this method, we choose a clean sky-patch located in the northern sky. The map-depth are calculated from a reasonable hit-map shown in Figure \ref{fig:hitmap}. We compare the reconstructed CMB signals and foreground residuals at the map level, the angular power spectra level, and the final $r$ constraints level using $500$ independent end-to-end simulations. The results are shown in Figure \ref{fig:CNILC_result} and Figure \ref{fig:rlimits}. Moreover, it is evident that the bias resulting from the foreground residuals is substantially reduced with the incorporation of additional prior information. The bias of the tensor-to-scalar ratio decreases from $0.05$ to $0.007$. However, this improvement comes at the expense of increased uncertainty, which rises from $0.025$ to $0.045$. In practice, it is necessary to strike a balance between these two aspects.

% In general, the limitations of the ILC method itself result in a decrease in its effectiveness for foreground removal at low signal-to-noise ratios. However, by introducing additional prior information and make it into a semi-blind method, the residual foreground can be significantly reduced. \tblue{In the current status where the tensor-to-scalar ratio has been constrained to $r < 0.036$ at 95\%~CL, further in-depth research on foreground phenomena is necessary to successfully distinguish the primordial B-modes from galactic foregrounds. This is because the complexity and intensity of foreground signals can introduce interference in the observation of the cosmic microwave background radiation. Additional foreground studies can involve detailed analysis of the physical properties, spatial distribution, and spectral characteristics of foreground components, as well as modeling and simulation of foregrounds.} Additionally, the output of the constrained NILC method is a sky map that preserves a great deal of information about the CMB, which is helpful for studying non-Gaussian CMB physics such as weak gravitational lensing and more. constrained NILC has the potential to serve as a viable method for mitigating foreground contamination in upcoming ground-based primordial gravitational wave experiments.

Furthermore, the output of the constrained NILC method, which is a sky map that retains a significant amount of information about the CMB, can be valuable for investigating non-Gaussian CMB physics such as weak gravitational lensing. As a result, constrained NILC has the potential to be a viable method for mitigating foreground contamination in future ground-based experiments focused on primordial gravitational waves.

The effectiveness of this method also relies on the accuracy of prior information regarding the foreground. To successfully distinguish the primordial $B$-modes from galactic foregrounds, it is crucial to conduct comprehensive research on foreground phenomena, especially considering the current constraint on the tensor-to-scalar ratio ($r < 0.036$ at 95\% confidence level). The complexity and intensity of foreground signals can introduce interference in foreground removal techniques. Further foreground studies should involve in-depth analysis of the physical properties, spatial distribution, and spectral characteristics of foreground components, along with modeling and simulation.

\acknowledgments
We thank Jacques Delabrouille, Mathieu Remazeilles, and those of AliCPT group for useful discussion. This study is supported in part by the National Key R\&D Program of China No.2020YFC2201600 and by the NSFC No.11653002.

\appendix

\section{The optimal EB-leakage correction}\label{sec:appendix}

EB-leakage arises when performing spherical harmonic transformation on a partial sky, as the orthogonality of spherical harmonics is no longer satisfied. In ground-based CMB observations, such leakage is unavoidable due to limited sky coverage, leading to bias on $r$. Correcting this EB-leakage is necessary.

Recent studies show that there's a good way called template cleaning method for a blind estimation and correction of $EB$ leakage \cite{PhysRevD.100.023538} comes from the partial sky, and can directly give a pure $B$ map, here we propose to extend the ILC method for scalar
$T$ map to $B$ map, as $B$ is a pseudo-scalar for a strictly speaking.

%\textcolor{blue}{functions can be demonstrated here.} 

\begin{figure}
    \centering
    \subfloat[]{\includegraphics[height=4.5cm]{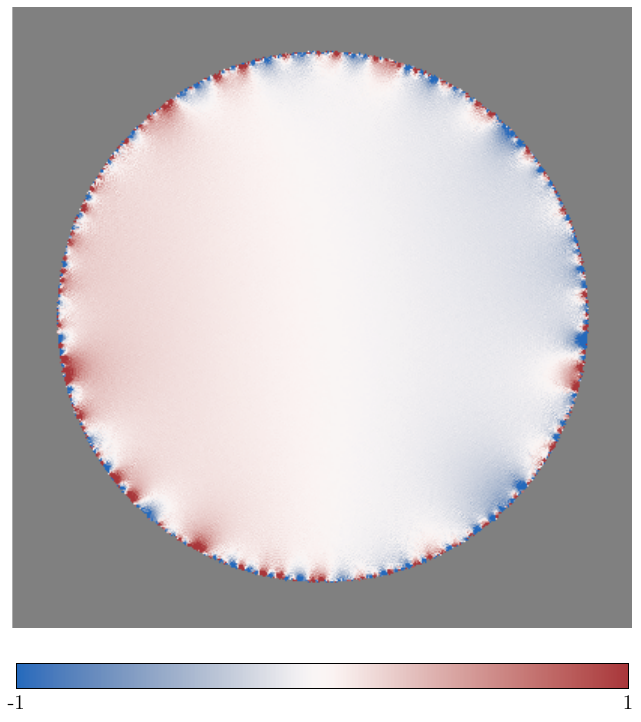}}
    \subfloat[]{\includegraphics[height=4.5cm]{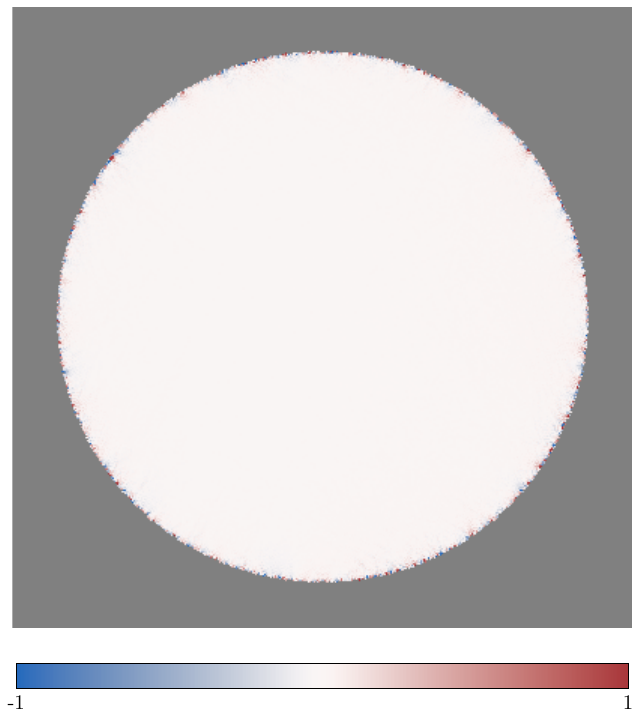}}
    \subfloat[]{\includegraphics[height=4.5cm]{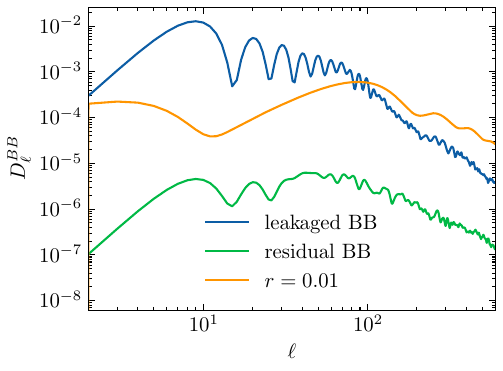}}
    \caption{(a): The leaked $B$ mode signal at the map level.
(b): The residual after $EB$ leakage correction using the method.
(c): The power spectra corresponding to (a) and (b), including the $r=0.01$ $B$-mode.}
    \label{fig:EBLeakResult}
\end{figure}

%Correction of $EB$ leakage on pixel domain is not computationally expensive. Briefly speaking, the procedure is:
The detailed description of the blind $EB$ correction method refers to paper \cite{PhysRevD.100.023538}, we briefly summarize the steps for pure $B$ map construction in the following:
\begin{enumerate}
    \item Use the \texttt{map2alm} function of \texttt{healpy} to directly transform the incomplete observation $(Q,U)$ maps to $(a_{\ell m}^{E},a_{\ell m}^{B})$. Then perform the inverse transformation from $(a_{\ell m}^{E}, 0)$ to $(Q, U)$ map, denoted as $(Q^{(1)}_E, U^{(1)}_E)$. Utilize $a_{\ell m}^{B}$ solely to obtain the pseudo-scalar $B$ map, denoted as $B^{(1)}$. At this stage, the $B$ map $B^{(1)}$ is contaminated by $EB$ leakage.
    \item Obtain the $B$ map from $(Q^{(1)}_E, U^{(1)}_E)$ using the same mask, denoted as $B^{(2)}$ in a similar manner.
    \item Now, $B^{(2)}$ serves as the leakage template for $B^{(1)}$ within the available region. The final leakage-corrected $B$ map can be expressed as:
\begin{equation}
  B_{\rm clean}(p) = B^{(1)}(p) - aB^{(2)}(p), \label{eq:ebleak}  
\end{equation}

Here, the factor $a$ is obtained through a linear fitting of $B^{(1)}$ and $B^{(2)}$, and $p$ denotes the pixel location.
\end{enumerate}
Figure \ref{fig:EBLeakResult} displays the results before and after $EB$ leakage correction using this method on a simple circular patch. The input sky-map has zero $B$ mode. It is evident that the $EB$ leakage resulting from the partial sky is greater than $r=0.01$ in the power spectrum, which heavily biases the $r$ measurement, particularly on the degree scale. After the correction, the residual leakage is much smaller than $r=0.01$. 

% \nocite{*}
\bibliography{main}
\bibliographystyle{JHEP} 

\end{document}